\documentclass[aps,prb,twocolumn]{revtex4}

\usepackage{graphicx}
\usepackage{dcolumn}
\usepackage{bm}

\newcommand{\beq}{\begin{equation}}
\newcommand{\eeq}{\end{equation}}
\newcommand{\abs}[1]{|#1|}

\begin{document}
\title{Absolute Rate Theories of Epigenetic Stability}
\author{Aleksandra M. Walczak$^*$}
\author{Jos\'{e} N. Onuchic$^*$}
\author{Peter G. Wolynes$^{*,\dagger}$} 
\affiliation{$^*$Department of Physics and Center for Theoretical Biological Physics,
  $^{\dagger}$Department of Chemistry and Biochemistry\\
University of California at San Diego, La Jolla, CA 92093, USA \\}
\date{\today}
\begin{abstract}
Spontaneous switching events in most characterized genetic switches are rare, resulting in extremely stable epigenetic properties. We show how simple arguments lead to theories of the rate of such events much like the absolute rate theory of chemical reactions corrected by a transmission factor. Both the probability of the rare cellular states that allow epigenetic escape, and the transmission factor, depend on the rates of DNA binding and unbinding events and on the rates of protein synthesis and degradation. Different mechanisms of escape from the stable attractors occur in the nonadiabatic, weakly adiabatic and strictly adiabatic regimes, characterized by the relative values of those input rates.
\end{abstract}
\maketitle
Information may be passed from one cellular generation to another not just in the form of the DNA sequence, but also in the long lived expression patterns of genes. The epigenetic state of the cell i.e. which genes are expressed at a given time, is determined in part by binding and unbinding of transcription factor proteins to the DNA. The genes with their partner proteins form complex dynamical systems known as genetic networks, which can have many steady states i.e. an attractor landscape \cite{Kauffman, SasaiWolynes}. The attractors are more stable than the individual molecular protein-DNA adducts, because the proteomic atmosphere of gene products renews the DNAs binding state, ultimately creating auto-catalytically its own proteomic atmosphere \cite{MonodSavageau, Ptashne, SasaiWolynes}. The attractors of such a genetic network may be associated with distinct cell types \cite{Kauffman, Davidson}. Experimental evidence for this view has recently been presented \cite{BarYam, ABOuden}. The growing experimental interest in this problem \cite{ABOuden}, as well as a number of theoretical puzzles involving the stability of the attractors \cite{AurellSneppen,Little}, call for a flexible and intuitive theory of the lifetime of such genetic network attractors. Some progress has already been made towards the goal \cite{WarrentenWolde, Bialek, AurellSneppen,MetzlerWolynes}, but existing formalisms are cumbersome, certainly when compared with the theory of activated events in molecular systems based ultimately on transition state ideas \cite{OnuchicWolynes88, WolynesSF}. Our goal here is to present a simple treatment of the noise induced transitions between two attractors on a landscape that is parallel to the treatment of simple molecular rate processes, which starts with Wigner's absolute rate theory \cite{Wigner}. In chemical kinetics, the ratio of escape is proportional to the probability of rare configurations equally likely to become reactant or product. These rare configurations represent a stochastic separatrix of the motion.\\
While thermal atomic motions cause the escape from energy minima in molecular physics, the noise in genetic networks comes from the probabilistic nature of the chemical reactions, since only a small number of proteins and individual copies of the target DNA are involved. Unlike molecules, genetic systems being far from equilibrium, cannot strictly be described by thermodynamic free energy functions. 
The stochastic separatrix for molecular activated events is a dividing surface passing through saddle regions of the free energy. We argue that, even in the absence of a free energy function, the notion of a stochastic separatrix between basins of attraction remains a good approximation \cite{WarrentenWolde, MetzlerWolynes} and allows a treatment of stochastic switching along the lines of a transition state theory with dynamic corrections involving the rates of elementary processes \cite{FrauenfelderWolynes, OnuchicWolynes88}.\\
\begin{figure}
\includegraphics[scale=0.45]{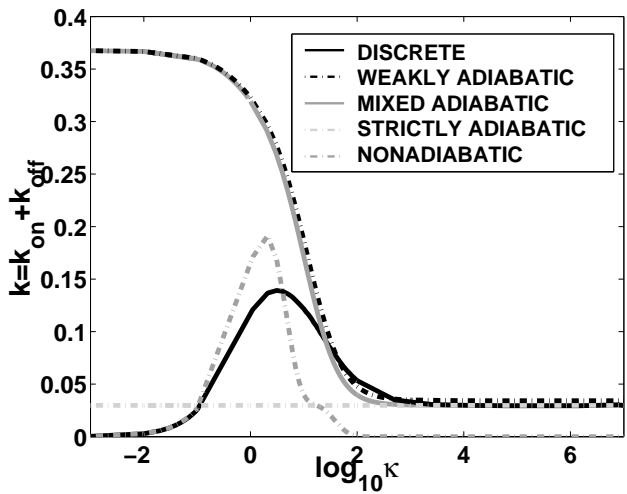}
\caption{The sum of the escape rates $k=k_{on}+k_{off}$ as a function of the adiabiaticity parameter $\kappa=\frac{h g_{\uparrow}^2}{2 k^3}$ for a self-activating switch with $g_{\uparrow}=100,g_{\downarrow}=8,k=1,n^{\dagger}_N=53.4$.  Comparison of the exact discrete $n$ numerical calculation based on the mean free passage time (black solid line), with approximate methods: in the nonadiabatic limit (small $\kappa$) (gray dashed line, Eqs \ref{kon_0} and \ref{koff_0}), in the weakly adiabatic regime (black dashed line, Eqs \ref{kon_adiab} and \ref{koff_adiab}) and mixed crossover regime (gray solid line). The adiabatic results tend asymptotically to the strictly adiabatic limit (large $\kappa$-flat escape rate) (light gray dashed line, Eqs \ref{kon_STRadiab} and \ref{koff_adiab}). In the strictly adiabatic limit the binding of transcription factors to the DNA binding site is equilibrated. In the nonadiabatic and weakly adiabatic limits the escape rates show a dependence on the adiabaticity parameter- the process in influenced by the DNA binding state fluctuations.}
\label{Figure1}
\end{figure}
The dynamics of gene networks involves two very different processes whose rates must be compared- protein synthesis and DNA binding. The complexity and energy consuming nature of protein synthesis, in prokaryotic cells, generally causes changes in protein number to take longer than the diffusion controlled binding times of transcription factors, even at their low concentrations. For this reason, it has been argued that one can describe the binding of the transcription factors to each DNA binding site as an instantaneously equilibrated process, when considering protein production. For steady states this approximation appears to be reasonably accurate. It has, however, also been noted \cite{Pedraza, Paulsson, Walczak, Hornos}, that the DNA state fluctuations may influence the protein number state fluctuations. Here we show that the impact of DNA state fluctuations on the escape process, is considerable in a rather wide parameter regime for which the steady states are not much influenced by the DNA state fluctuations (Figure \ref{Figure1}). For biologically relevant parameters, the DNA occupancy fluctuations may significantly increase the spontaneous switching rate from a given attractor. In what we call the nonadiabatic limit, where DNA state fluctuations dominate the protein number fluctuations, individual binding and unbinding events of the transcription factors are directly responsible for the transition. For much of the adiabatic regime, although the influence of DNA fluctuations on the steady state protein levels is negligible, these fluctuations still modify the lifetime of a state - we will call these transitions "weakly adiabatic". DNA occupancy fluctuations can only be neglected at very high values of the rate ratios, in what we call the strongly adiabatic limit.\\
As Sasai and Wolynes \cite{SasaiWolynes} have pointed out, the stochastic theory of a simple genetic switch, can be considered analogous to the physicists' Kondo problem or the chemists' electron transfer process, where DNA occupancy plays the role of a spin or electronic state variable \cite{OnuchicWolynes88}. In a simple and intuitive way, here we exploit these analogies to compute the lifetime of a genetic switch, using the idea of a landscape with a stochastic separatrix \cite{WarrentenWolde}, much like in the earlier proposed threshold model \cite{MetzlerWolynes}. Our treatment is quite analogous to that used for characterizing adiabatic vs nonadiabatic regimes of quantum rates \cite{OnuchicWolynes88, WolynesSF}. The approach, we present is easily generalizable to a many gene system. In the general case the present approximation yields the lifetime of a given state of the switch, which is governed purely by a few local properties of the landscape and does not require computing complicated trajectories. Global properties, sensed by the most probable escape paths, generally enter rates for far-from-equilibrium systems \cite{SteinMaier}. The present approach must, therefore, be admitted to be approximate. The simplicity hopefully will make up for some inaccuracy.\\
Several treatments of the mean first passage time between epigenetic states have already appeared. Most of these studies assume the DNA state equilibrates on a much faster time scale than the protein number \cite{KeplerElston,Romaetal}. We refer to this as the adiabatic limit. In this limit the protein number states may then be treated as a continous variable giving an expression for the mean first passage time \`a la the Smoluchowski theory of diffusive rates as sketched by Bialek \cite{Bialek}. A more rigorous approach finds the rate by constructing the most probable escape path \cite{AurellSneppen} or by calculating the distribution of paths \cite{WarrentenWolde, AllenWarrentenWolde}. These methods are powerful, but they are hard to visualize, especially for more complex switching systems. While the usually invoked adiabatic limit seems to be appropriate for simple switches in prokaryotes, it is not an obviously correct approximation for switches that have more complex operators, in which multiple protein elements must combinatorically assemble at a given site, slowing the binding \cite{Hwa}. Nonadiabatic effects should also play a significant role in eukaryotic systems where chromosome restructuring, which may be quite slow, dominates the epigenetic transition. Artificially engineered switches \cite{OzbudakOuden} may be constructed with parameters spanning the entire phase diagram.\\
\textbf{The Simplest Switch}\\
For illustration we will present our ideas using the simplest example of a system in which we can consider the escape from one minimum to another - a bistable self-activating switch \cite{Hornos}. We emphasize the approach is more generally applicable. The self-activating switch consists of a single gene, which may be found in one of two states: on or off. In the off state proteins are produced at a basal level, but in the  on state proteins are produced at an enhanced level, leading to a number, $n$, of proteins in the cell at any moment. The proteins act as activators by binding to the same operator site as the gene governing their production. We assume they bind as dimers with a rate $h(n)=hn(n-1)/2$. The unbinding of the transcription factors is described by a rate $f$. We neglect time delays due to mRNA synthesis etc. (which admittedly may play a key role), so that protein population dynamics is governed by a birth death process. Protein degradation occurs with a rate $k$, production with the activated rate $g_{\uparrow}$ in the on state and the basal rate $g_{\downarrow}$ in the off state. The system is characterized by a two state joint probability distribution $\vec{P}(n)$, describing the probability of having $n$ proteins in the system and the DNA binding site being in the bound (on-$\uparrow$) or unbound (off-$\downarrow$) state. A recent combined experimental and theoretical study \cite{OzbudakOuden} has brought attention to the bistability of a switch in previously unexplored limits, when the degree of operon repression is small. Our discussion will turn also to the nonadiabatic limit. Here the equilibration of the DNA and changes in the protein number occur on comparable time scales. \\
To compute escape rates from the steady state attractors one must determine the stochastic separatrix \cite{WarrentenWolde}. In the adiabatic limit, the position $n^{\dagger}_A$ of the minimum of the total probability distribution $P(n)=P_{\uparrow}(n)+P_{\downarrow}(n)$ is given by the condition of zero mean protein flow $dn/dt |_{|_{n=n^{\dagger}_A}}=(fg_{\downarrow}+h(n)g_{\uparrow})/(f+h(n))-kn |_{n=n^{\dagger}_A}=0$. For a bistable switch, this equation is satisfied by three values of $n$; - one solution gives the separatrix, the other two the positions of the high and low protein number stable steady state attractors, $n^{\downarrow}_A$ and $n^{\uparrow}_A$. In the nonadiabatic limit the stochastic separatrix refers both to the DNA and protein number state. This results in a different value of the critical separatrix numbers $n^{\dagger}_N$ in the nonadiabatic, and $n^{\dagger}_A$ in the adiabatic limits. The direction of flow changes when the DNA state changes. Therefore the position of $n^{\dagger}_N$ corresponds to that number of proteins needed for the system to have comparable probability to be in the on or the off state. For simplicity we can approximate in the large $n$ limit $h(n)=h/2n(n-1)\approx hn^2/2$ and determine the position of the nonadiabatic separatrix by means of mass action, using the chemical equilibrium constant $K^{eq}$: $n^{\dagger}_N=V \sqrt{K^{eq}}$, where $K^{eq}=2f/(hV^2)$, where $V$ is the cell volume. The steady state attractors in the nonadiabatic limit are determined by the birth-death processes in the particular DNA states: $n^{\downarrow}=g_{\downarrow}/k$ in the off state and $n^{\uparrow}=g_{\uparrow}/k$ in the on state. To function as a switch $n^{\downarrow}$ must be less than $n^{\dagger}_N$ and $n^{\uparrow}$ must be greater than $n^{\dagger}_N$. We can rewrite the adiabatic separatrix positions in terms of the volume scaled equilibrium constant $K^{eq}V^2$, which scales with $n^{\dagger 2}_N$, as $n^{\dagger}_A= K^{eq}/n^{\uparrow}$ and $n^{\downarrow}<n^{\downarrow}_A<n^{\dagger}_A<n^{\dagger}_N<n^{\uparrow}_A<n^{\uparrow}$.\\
\textbf{Nonadiabatic Rate Theory}\\
Here we compute the rate of escape of the system from the low protein number attractor to the high protein number attractor ($k_{on}$) and vice versa  ($k_{off}$). Since in the nonadiabatic limit the low protein number attractor corresponds to the off DNA occupancy state and the high protein number state corresponds to the on DNA occupancy state, the transition from the low protein number state to the high protein number state is requires the binding of an activator. Without the possibility of binding and unbinding, the dynamics in each attractor would be described by stochastic destruction and production of proteins alone, resulting in fluctuations of the mean protein number around each steady state. Consider a system maintained in the off DNA binding state and that now has $n^{\downarrow}$ proteins. The initial probability of being in the off DNA state, with precisely $n^{\downarrow}$ proteins present is $p_{off}(n^{\downarrow})=P_{\downarrow}(n^{\downarrow})/(P_{\uparrow}(n^{\downarrow})+P_{\downarrow}(n^{\downarrow}))$. $n^{\downarrow}$ may be generally assumed to be close to the mean number of proteins in the off state ($n^{\downarrow}=g_{\downarrow}/k$). If a binding event now occurs at time $t=0$, the gene spontaneously flips into the on state and proteins are now produced at an enhanced rate. The protein number increases towards the mean number in the high protein state ($n^{\uparrow}=g_{\uparrow}/k$). If the activator does not unbind before the number of proteins becomes characteristic of the on state attractor a successful switching event will have taken place and the protein number will now fluctuate around the on steady state value. However, since, we are in the nonadiabatic limit, the timescales to reach the steady state for both the DNA binding state and protein synthesis and degradation are assumed comparable, so an activator may in fact unbind before reaching the separatrix at $n^{\dagger}_N$. If an activator does unbind during that time, the gene returns to an off state, albeit with a slightly higher number of proteins than initially. Another binding event will repeat the above scenario, until the protein number safely crosses the separatrix at $n^{\dagger}_N$ and the steady state corresponding to an activated gene is reached (Figure \ref{Picture} $a$). The average time needed to cross the barrier from an initial point $n^{\downarrow}$, which is also the time allowed for a unbinding event to occur, is the mean time to reach $n^{\dagger}_N$ for the enhanced production rate. The initial rate of binding an activator $h(n^{\downarrow})=h/2 n^{\downarrow}(n^{\downarrow}-1)$ must be modified to account for the possibility of unbinding again before the system crosses the separatrix. Summing of these attempted crossings, results in an expression for the rate of escape from the off state minimum ($n^{\downarrow}$) to the on state ($n>n^{\dagger}_N$) in the nonadiabatic regime given by:
\begin{equation}
k_{on}(n^{\downarrow})= p_{off}(n^{\downarrow}) h(n^{\downarrow})e^{-\int_{t(n^{\downarrow})}^{t(n^{\dagger}_N)} f dt}
\label{kon_0}
\end{equation}
The exponential term gives the successful fraction of attempts to reach the protein number based separatrix, launched from the steady state $n^{\downarrow}$. The total time to reach the separatrix is given by $t(n^{\dagger}_N)-t(n^{\downarrow})$, as determined by the average flows in the initial DNA state and the mean time for an unbinding event to occur is $f^{-1}$. Explicitly, the escape rate from the off state, becomes $k_{on}(n^{\downarrow})= p_{off}(n^{\downarrow}) h(n^{\downarrow})(({g_{\uparrow}-kn^{\downarrow})/(g_{\uparrow}-kn^{\dagger}_N}))^{-f/k}$. The power-law term describes the motion on the surface with enhanced production after binding of the activator. In the nonadiabatic limit, the probability distributions for the on and off states are unimodal. Therefore it is unlikely for the gene to be in the on state is the number of proteins is small, thus $p_{off}(n^{\downarrow})\approx 1$. If the protein number is large and the unbinding rate is comparable to the death rate this expression yields:\\
\begin{equation}
k_{on}(n^{\downarrow}) \sim h(n^{\downarrow}) e^{-\frac{f}{k}(n^{\dagger}_N-n^{\downarrow})} \sim h(n^{\downarrow}) e^{- \kappa \frac{\sqrt{(K^{eq}V^2)^3}}{{n^{\uparrow }}^2}}
\label{kon}
\end{equation}
where $\kappa=h g_{\uparrow}^2/(2 k^3)$. In the extreme nonadiabatic limit $\kappa\rightarrow 0$, the first attempt may be successful hence the result simplifies to $k_{on}(n^{\downarrow})\sim h(n^{\downarrow})$.\\
A similar calculation can be carried out starting from the other steady state. The escape rate from the on state, with $n^{\uparrow}>n^{\dagger}_N$ proteins, is given by the rate of binding of an activator at time $t=0$, providing the system is in the on state $p_{on}(n^{\uparrow})=P_{\uparrow}(n^{\uparrow})/(P_{\uparrow}(n^{\uparrow})+P_{\downarrow}(n^{\uparrow}))$, reduced by the probability that an activator rebinds before the protein number decreases to numbers characteristic of in the off state ($n<n^{\dagger}_N$). The time available to rebind is calculated using protein production at a basal level. The $k_{off}$ rate is therefore:
\begin{equation}
k_{off}(n^{\uparrow})=p_{on}(n^{\uparrow})  f e^{{-\int_{t(n^{\uparrow})}^{t(n^{\dagger}_N)} h[n(t)] dt}}
\label{koff_0}
\end{equation} 
For the off rate the mean free path before a rebinding event depends on the mean number of proteins in the system $n$. The argument of the exponential still describes the number of rebinding events. In the strongly nonadiabatic case, $p_{on}(n^{\uparrow}) \approx 1$, and for very large mean protein numbers the escape rate tends to:\\
\begin{equation}
k_{off}(n^{\uparrow}) \sim f e^{-\frac{h}{4k}({n^{\uparrow}}^2-{n^{\dagger}_N}^2)}\sim f e^{-\frac{\kappa}{2}\frac{{n^{\uparrow}}^2-K^{eq}V^2}{{n^{\uparrow}}^2}}
\label{koff}
\end{equation} 
Due to the timescale separation in the nonadiabatic limit the system may be approximated as a two state system. The ratio of the escape rates, therefore yields the ratio of the probabilities to be in the individual steady states. The equilibrium constant for the "dressed" genetic states in the nonadiabatic limit ${K}^{GS}=k_{off}/k_{on}$ therefore becomes ${K}^{GS}\approx (n^{\dagger}_N/n^{\downarrow})^2 exp(-\kappa/2)=K^{eq}V^2/{n^{\downarrow}}^2 exp(-\kappa/2)$. When $\kappa=0$ the proteomic atmosphere has no effect on the relative stability of the DNA occupancy, which follows the ordinary mass action law.\\
The formulae described above provide quite intuitive representations of specific escape mechanisms. These results may also be formally obtained via the path integral solution of the master equation by using the method described by Wang, Onuchic and Wolynes \cite{WOW} for kinetic protein folding. This result also coincides with the heuristic approach of Ninio \cite{Ninio}.\\
\textbf{Adiabatic Rate Theories: Weak and Strong Regimes}\\
In the nonadiabatic limit the switch reaches the separatrix within the time for a few binding events, as schematically portrayed in panel $a$ of Figure \ref{Picture}. In what we call the weakly adiabatic regime, the escape process proceeds differently. The DNA occupancy responds quickly to the changing proteomic atmosphere reaching a local steady state before the protein number changes by a large amount. The average occupancy then determines the average local rate of protein synthesis and degradation. A few binding and unbinding events are required in the nonadiabatic limit, but in the adiabatic limit those events are much too common to allow the direct mechanism. One is tempted to equate the local diffusion rate to that coming from synthesis and degradation. But this temptation can only be rigorously indulged at an extraordinary high binding rate. Instead a random, but cyclic process of binding, growth and unbinding churns the protein number like a turbulent surf. The cyclic motions of eddies in an ocean wave, if interrupted contribute to a diffusive transport of flotsam to the shore. In the same way, in most of the weak adiabatic regime, protein numbers fluctuate from the mean flow through this "churning mechanism". The protein number, changes slightly with each cycle of binding/growth/unbinding and eventually reaches the separatrix point due to the resulting diffusive motion. One can show the system acts as if it were diffusing along an effective potential, whose gradient gives the mean flow expected from the average occupancy $V(n)=g_{eff}(n)-kn$ (panel $b$ in Figure \ref{Picture}). The diffusion rate in this outwardly adiabatic regime though depends on the nonadiabatic events. Only at very high adiabaticity is diffusion ascribable to birth-death alone. \\
\begin{figure}
\includegraphics[height=7.5cm,width=8.5cm]{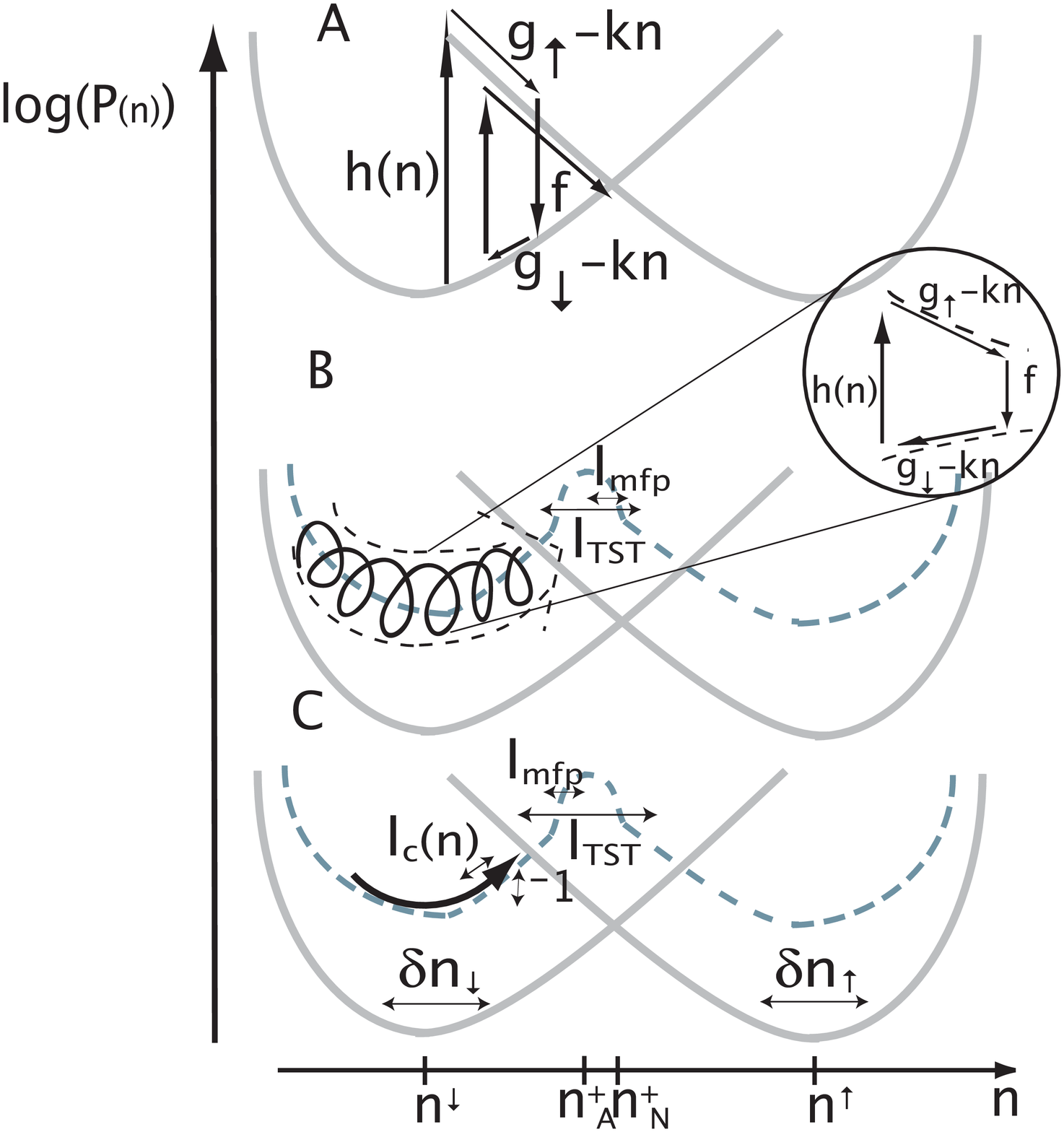}
\caption{A schematic diagram of the difference in the character of the transitions from the state with a small number of proteins to the state with a large mean steady state number of proteins in the nonadiabatic (A) where the escape rate is given by Eqs \ref{kon_0} and \ref{koff_0}, adiabatic (B),  where the escape rate is given by Eqs \ref{kon_adiab} and \ref{koff_adiab}, and extremely adiabatic (C) regimes. The dark gray line marks the effective potential for protein number change. The horizontal arrows signify binding and unbinding events.}
\label{Picture}
\end{figure}
It is helpful to understand the "eddy-induced" diffusion in an intuitive way. The effective production rate $g_{eff}(n)=(fg_{\downarrow}+h(n)g_{\uparrow})/(f+h(n))$ is the production rate averaged over the binding and unbinding states, if they were in equilibrium. The diffusion expected solely from the birth-death processes would just be $D_{BD}(n)=g_{eff}(n)+kn$. This fluctuation mechanism is augmented by diffusion in the orthogonal two state "binding-space", that is the eddy motion. The difference in the mean distance in protein number that would be travelled in the two DNA states during a typical eddy cycle will be $\Delta n=(g_{\uparrow}-g_{\downarrow})/(f+h(n))$. It is the typical difference in protein number expected after a full cycle of an eddy has been traversed. It is given by the difference in velocity in protein number space in a given binding state, $\Delta v=\abs{g_{\uparrow}-g_{\downarrow}}$, times the mean time before the binding state changes $\Delta t=(f+h(n))^{-1}$, such that $\Delta n=\Delta v \Delta t$. The mean free time, or the eddy mixing time, is given by the sum of the characteristic times for binding and unbinding, both of which must occur to return to the original binding state, $\tau={f}^{-1}+({h(n)})^{-1}$. The rate which describes the eddy cycling thus becomes $\tau^{-1}=fh(n)/(f+h(n))$. The diffusion coefficient $D=\Delta n^2/\tau$ is the square of the mean change in protein number divided by the characteristic time spent within a given eddy. The latter depends on both binding and unbinding events. One thus finds in $D_{binding}(n)=fh(n)(g_{\uparrow}-g_{\downarrow})^2/(f+h(n))^3$. \\
The mean number of proteins of a given type produced in the active state is of the order of $g_{\uparrow}/k\sim10^2$. The degradation rate of proteins gives lifetimes of the order of a bacterial generation $k\sim10^{-3} s^{-1}$. Dissociation rates from the DNA vary from $f \sim1-10^{-3}s^{-1}$ and typical equilibrium constants may be taken to be $K^{eq} V^2\sim10^2-10^4$, which results in association constants $h/2=f/(K^{eq}V^2)\sim10^{-2}-10^{-7}s^{-1}$ (based on $\lambda$ phage data as assembled in \cite{AurellSneppenPRE} and references therein). We therefore see that typical adiabaticity parameters scan a wide range: $\kappa=h g_{\uparrow}^2/(2 k^3)\sim 10^0-10^{5}$. The diffusion coefficient from churning, which depends on the DNA occupancy dynamics, typically influences the escape rate over four orders of magnitude of the adiabaticity parameter $\kappa \in (10^0-10^{4})$, nearly covering the biologically relevant regime. For escape processes the DNA binding dynamics cannot be neglected until the adiabaticity parameter becomes extremely large ultimately yielding the strongly adiabatic regime. As shown in Figure \ref{Picture}, the eddies due to the influence of the DNA binding state become smaller with faster binding, until the motion becomes dominated by simple birth-death diffusion along the effective potential, giving the steady state probabilities, averaged over the DNA binding states (panel $c$ of  Figure \ref{Picture}).\\
In the adiabatic limit, the escape rate is governed largely by the fraction of systems at the separatrix $N^{\dagger}_A$ compared to the fraction residing near the original attractor $N_{attr}$: ${N^{\dagger}_A}/{N_{attr}}={P(n^{\dagger}_A)}/{p^{s}_< (n^{\dagger}_A)}$, where $p^{s}_< (n^{\dagger}_A)=\sum_{n<n^{\dagger}_A} P(n)$ and $P(n)$ is the steady state probability density for a state with $n$ proteins. Clearly $p^{s}_< (n^{\dagger}_A)= P(n_{in}) \delta n_{in}$, where $\delta n_{in}$ is the width of the attractor. It is important to understand the spatial variation of $P(n)$, described by the "potentials" in Fig. \ref{Picture}. The spatial variation depends on the balance of the local mean flow against the flow due to diffusion. We can understand this balance by considering the motion pictured in Figure \ref{Picture} $b$. The mean local velocity by which the protein number changes is $\bar{v}=g_{eff}-kn$. In addition to this drift the protein number changes by diffusive motion from places of low to high probability, with a velocity of $v_{diffusion}=2D^i(n)/l_{c}$, where $l_c$ is a characteristic "lengthscale" over which the steady state probability changes by roughly one e-fold. $D^{i}(n)$ refers to the diffusion coefficient, which governs the motion in a particular regime. It is equal to $D_{binding}(n)$ in the weakly adiabatic regime, $D_{BD}(n)$ in the strictly adiabatic regime and is roughly $D_{BD}(n)+D_{binding}(n)$ in the small crossover region in between. To traverse this scale the local velocity has to be at least as large as the velocity of the diffusive motion $\bar{v} \geq v_{diffusion}$. The equality $\bar{v} =v_{diffusion}$ sets a characteristic length scale of the problem $l_c=2 D^i(n)/\abs{\bar{v}(n)}$, over which locally the probability in a steady state should change by a factor of $e$. This relation is valid both in the adiabatic and nonadiabatic regimes. The quantity $l_c$ is analogous to the "scale height" in the equilibrium barometric problem. How many of these characteristic steps of length $l_c$ are needed in order for the system to reach $n^{\dagger}_A$ from its steady state value? Bearing in mind that the length of each step, depends on $n$, we must concatenate these steps to give the probability to be at the separatrix relative to being near the initial state. The probability exponentially depends on the number of scale heights of varying length $l_c$ needed to reach the improbable separatrix starting from the most probable situation at the basin center, $\exp{[-2 \int_{n^{\downarrow}_A}^{n^{\dagger}_A} dn l_c^{-1}]}$.\\
To find the rate, we finally need the width $\delta n_{in}$. The size of the attractor $\delta n_{in}$ is analogous to $l_c$ at the bottom of the basin, but quadratic order effects must be included. To compare the velocities of the motion near the basin center due to drift and diffusion, the drift velocity must be computed as the "drift frequency" in the initial state $\omega(n_{in})=(\partial v(n)/\partial n)|_{n_{in}}=fhn_{in}(g_{\uparrow}-g_{\downarrow})/(f+h(n_{in}))^2-k$ times the distance from the stationary point. Comparing drift and diffusion velocities in the same region $\omega(n_{in}) \delta n_{in}=D^i(n_{in})/\delta n_{in}$, gives the size of the attractor $\delta n_{in}=\sqrt{\abs{D^i(n_{in})/\omega(n_{in})}}$. The exponential term counts the paths from all possible position within the attractor. We must therefore divide the by the width of the attractor.  \\
To determine the epigenetic escape rate we need also the transmission factor. In the adiabatic limit, reaching the separatrix does not yet guarantee a successful escape. Once the protein number reaches the vicinity of the stochastic separatrix the system may directly cross the separatrix, or recross it many times before committing to the new attractor. The number of escapes per unit time rate is thus proportional to the velocity with which the system moves over the separatrix, divided by the number of attempts before it successfully commits to the new attractor $k=\delta v/N P(n_{in}\rightarrow n\sim peak)$. The velocity around the peak is determined by a mean free path for number fluctuations $l_{mfp}$ and a mean free time $\tau$ relevant to that region, $\delta v=l_{mfp}/\tau$. Only in the crossover region is it necessary to take all processes into account on equal footing when evaluating the mean free path $l_{mfp}$ and the associated mean free time $\tau$. In the weakly and strongly adiabatic limits the results simplify. In the weakly adiabatic region, the mean free path is dominated by the DNA churning cycles and is given by the typical eddy size $l_{mfp}\approx (g_{\uparrow}-g_{\downarrow})/(f+h(n))$ and $\tau={f}^{-1}+({h(n))^{-1}}$. In the strictly adiabatic limit, the motion is determined by the birth and death of proteins. Effectively, the protein number changes by $l_{mfp}$ equal to one protein in the mean free time $\tau=(g_{eff}(n)+k)^{-1}$. Once the mean free path has been determined, the number of crossings is then the number of steps of the size of the mean free path needed to cross the transition state region $l_{TST}$. Like the basin size, the size of the transition state region is $l_{TST}=\sqrt{D^i(n^{\dagger}_A)/\omega(n^{\dagger}_A)}$. The escape rate from the left attractor, $n^{\downarrow}_A$, is $k_{on}(n^{\downarrow}_A)=l^2_{mfp}/(l_{TST} \tau) ({\delta n^{\downarrow}_A})^{-1} e^{-\int_{n^{\downarrow}_A}^{n^{\dagger}_A} dn l_c^{-1}}$, where $l^2_{mfp}/\tau=D_{i}(n^{\dagger}_A)$ and $i$ indicates $BD$, $binding$ and $mixed$ in the appropriate regimes. This gives the rate of the escape from the low protein number state in the weakly adiabatic regime: 
\begin{equation}
k_{on}(n^{\downarrow}_A)= \frac{k}{2 \pi} D^i(n^{\dagger}_A)\sqrt{\frac{\abs{\omega(n^{\downarrow}_A) \omega(n^{\dagger}_A)}}{D^i(n^{\dagger}_A)D^i(n^{\downarrow}_A)}} e^{ -2 \int_{n^{\downarrow}_A}^{n^{\dagger}_A} dn l_c^{-1}}
\label{kon_adiab}
\end{equation}
where $n^{\dagger}_A$ is the number of proteins corresponding to the minimum of the total steady state probability distribution. In the adiabatic regime the separatrix is given as the fixed point of the average flow: $g_{eff}(n^{\dagger}_A)=k n^{\dagger}_A$. \\
In the strictly adiabatic limit, the eddy motion may be neglected. So $l_c^{\kappa \rightarrow \infty}$ is determined solely by the equilibrated diffusion in protein number space $l_c^{\kappa \rightarrow \infty}\approx (g_{eff}+kn)/(g_{eff}-kn)$ . All the components in Eq \ref{kon_adiab} can be obtained using quadrature, in this case, yielding a complex expansion. A more simplified result, explicit in terms of chemical rate constants, follows if we linearize $l_c^{-1}$ in the region, which contributes most to the result of the integral. In this situation equation \ref{kon_adiab} becomes:
\begin{equation}
k_{on}(n^{\downarrow}_A)=\tilde{f_1}(K^{eq}) e^{-\frac{\abs{l_c^{-1}(n_{min})}}{2(n^{\dagger}_A-n_{min})}(n^{\dagger}_A-n^{\downarrow}_A)^2}
\label{kon_STRadiab}
\end{equation}
where $n_{min}$ number of proteins for which $l_c^{-1}$ has the largest value.  The largest value of $l_c^{-1}$ corresponds to the the smallest characteristic length scale in the region of integration. The value of $l_c^{-1}(n_{min})$ scales as $n_{min}\sim V\sqrt{K^{eq}/2}$. The pre-exponential factor has the form $\tilde{f_1}(K^{eq})=kV/(2\pi) \sqrt{ K^{eq}/(a_0 {n^{\uparrow}}^6)({n^{\uparrow}}^4-(K^{eq}V^2)^2-2K^{eq}V^2 (n^{\uparrow})^2)}$, where $a_0=g_{\downarrow}/g_{\uparrow}$. The escape rate decreases with the equilibrium constant and system size. Using the dependence of the minimum of the integrand as a function of the equilibrium constant $K^{eq}V^2$, one finds the escape rate scales as $e^{-\alpha_1{n^{\uparrow}}^{-2}(K^{eq}V^2-3a_0{n^{\uparrow}}^2)^{3/2}}$, where $\alpha_1$ is a numerical factor of the order of $1/2$. The rate of escaping from the off state attractor exponentially decreases with increasing of the equilibrium constant. \\ 
How the escape rate depends on the molecular parameters, can be seen by assuming, for simplicity, a highly cooperative variation of the equilibrium DNA occupancy with protein concentration. In this case the effective production rate can be approximated by the production rate in the off state attractor, $g^{eff}(n)\approx g_{\downarrow}$. Now, the protein dynamics will be determined by the rates characteristic of the attractors, until the system reaches the separatrix. This approximation is like the threshold picture of Metzler and Wolynes \cite{MetzlerWolynes}. In this approximation one finds:
\begin{equation}
k_{on}(n^{\downarrow}_A)=\tilde{f_1}(K^{eq}) e^{-\frac{1}{2} \frac{k(n^{\dagger}_A-n^{\downarrow}_A)^2}{kn^{\dagger}_A+g_{\downarrow}}}  
\label{kon_ROUGH}
\end{equation}
When the cell is sufficiently small the separatrix merges with both attractors. In such a regime, this simple formula correctly predicts the functional dependence of the escape rate on the equilibrium constant and the protein production rates. When the separatrix begins to merge the attractor, the exponential term approaches unity. Thus stability is compromised. When the attractors merge with the separatrix the pre-exponential factor becomes important for quantitative analysis \cite{Romaetal, AurellSneppen}.\\
In the $\kappa$ dependent weakly adiabatic region, the probability distributions look qualitatively similar to those in the strictly adiabatic limit: the extrema do not change as $\kappa$ increases. In the escape rate calculation, however, one compares the ratios of the probabilities near the minimum and the saddle regions. This ratio is significantly different in the weak and strong adiabatic regimes and strongly affects the spontaneous switching rates, as seen in Figure \ref{Figure1}. In the weak adiabatic regime one finds the escape rates depend exponentially on the adiabaticity parameter $\kappa$. The escape rate therefore is approximately dominated by $kV/(2 \pi) K^{eq}{n^{\uparrow}}^3/a_0 \sqrt{{n^{\uparrow}}^2-2 K^{eq} V^2}/({n^{\uparrow}}^2+K^{eq}V^2)^3\exp(-f/k K^{eq}V^2 a_0/n^{\uparrow} (n^{\dagger}_A-n^{\downarrow}_A)/(n^{\downarrow}_A n^{\dagger}_A))$. In the weakly adiabatic regime, the effective growth rate can be well approximated as that with a fixed DNA occupancy, as in the Metzler-Wolynes threshhold model \cite{MetzlerWolynes}.\\
\begin{figure}
\includegraphics[scale=0.4]{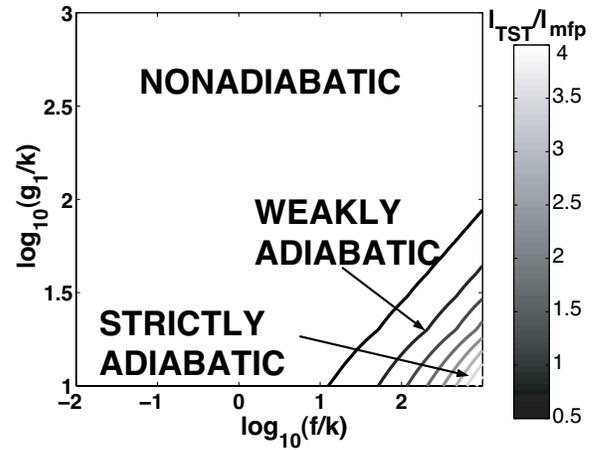}
\caption{A phase diagram as a function of the activated production rate $g_{\uparrow}$ and the unbinding rate $f$ for constant $K^{eq} V^2$, showing the areas of parameter space where a given escape mechanism dominates based on the ratio of the size of the transition state region $l_{TST}$ to the mean free path $l_{mfp}$. If the number of crossings of the separatrix is large $\frac{l_{TST}}{l_{mfp}}>1$, the transition is adiabatic. If the system commits to a new attractor after one crossing $\frac{l_{TST}}{l_{mfp}}<1$ the transition is nonadiabatic.}
\label{Diagram}
\end{figure}
The transition can be treated from the high protein number state to the low protein number state much as above in the adiabatic limit. The rate of escape from high protein number to the low protein number depends on the relative probability that the system is to the right of the separatrix, characterized by a mean protein number $n^{\uparrow}$ compared to the steady state probability of being at the separatrix $n^{\dagger}_A$, $k(n^{\uparrow}\rightarrow n\sim peak)=P(n^{\dagger}_A)/p^{s}_> (n^{\dagger}_A)$. $p^{s}_> (x)=\sum_{n=x}^{n=\infty} P(n)$. The escape rate turns out to be:
\begin{equation}
k_{off}(n^{\uparrow})= \frac{k}{2 \pi}D^i(n^{\dagger}_A)\sqrt{\frac{\abs{\omega(n^{\uparrow}) \omega(n^{\dagger}_A)}}{D^i(n^{\dagger}_A)D^i(n^{\uparrow})}} e^{-2 \int_{n^{\dagger}_A}^{n^{\uparrow}} dn l_c^{-1}}
\label{koff_adiab}
\end{equation}
We can approximate $l_c^{-1}$ in the strictly adiabatic limit as for the $k_{on}$ calculation. Then the strictly adiabatic escape rate becomes:
\begin{equation}
k_{off}(n^{\uparrow})=\tilde{f_2}(K^{eq}) e^{-\frac{l_c^{-1}(n_{max})}{4(n_{max}-n^{\dagger}_A)}(n^{\uparrow}_A-n^{\dagger}_A)^2}
\label{koff_STRadiab}
\end{equation}
where $n_{max}$ is the number of proteins at the maximum of $l_c^{-1}$, which scales as $n_{max}\sim =V\sqrt{K^{eq}}$. The pre-exponential fatcor has the form $\tilde{f_2}(K^{eq}) \approx kV/(2 \pi) ({n^{\uparrow}}^4-(K^{eq}V^2)^2-2 K^{eq} V^2{n^{\uparrow}}^2) \sqrt{K^{eq}}/{n^{\uparrow}}^5$. More explicitly the escape rate from the on state scales as $\sim e^{-\alpha_2 {n^{\uparrow}}^{-2}\sqrt{(\zeta {n^{\uparrow}}^2-K^{eq} V^2)}^{3}}$, where $\alpha_2\approx 2$ and $\zeta \approx 1/4+a_0/2$ are constant numerical factors. The escape rate from the on state attractor exponentially increases with the increase of the equilibrium constant. A simple result is also obtained by replacing the effective production rate by the value of the effective production rate in the on state attractor $g^{eff}(n)\approx g^{eff}(n^{\uparrow}_A)$:
\begin{equation}
k_{off}(n^{\uparrow}_A)= \tilde{f_2}(K^{eq})e^{-\frac{1}{2} \frac{k(n^{\uparrow}_A-n^{\dagger}_A)^2}{kn^{\dagger}_A+g^{eff}(n^{\uparrow}_A)}}   
\label{koff_ROUGH}
\end{equation}
The equilibrium constant for the dressed genetic switch state in the strongly adiabatic limit is $\bar{K}^{GS}=k_{off}/k_{on}\sim f_r(K^{eq}) e^{-{n^{\uparrow}}^{-3}(\beta_1 (\sqrt{(\zeta {n^{\uparrow}}^2-K^{eq}V^2)}^{3}-\beta_2 \sqrt{(K^{eq}V^2-3a_0{n^{\uparrow}}^2)}^{3})}$, which sharply depends on the proteomic atmosphere. $f_r(K^{eq})=\sqrt{a_0({n^{\uparrow}}^4-(K^{eq}V^2)^2-2 K^{eq}{n^{\uparrow}}^2)/{n^{\uparrow}}^4}$ $\beta_1\approx2$,  $\beta_2\approx1/4$ are numerical factors.\\
In the weakly adiabatic regime the exponential term in the off escape rate becomes $\exp(-h/(2k){n^{\uparrow}}^{-2}/(K^{eq}V^2)(1/6({n^{\dagger}_A})^{6}-({n^{\uparrow}_A})^{6}-g^{eff}(n^{\uparrow}_A)/(5k)(({n^{\dagger}_A})^{5}-({n^{\uparrow}_A})^{5})))$. So in the weakly adiabatic limit the equilibrium coefficient for the dressed genetic switch states $\bar{K}^{GS}=k_{off}/k_{on}$ scales as $\bar{K}^{GS}\sim a_0 {n^{\uparrow}}^3/\sqrt{K^{eq}V^2}^3  e^{-h/(2k) \xi_1(n^{\uparrow})^{-2}/(K^{eq}V^2)((n^{\uparrow})^{6}-\xi_2 (K^{eq}V^2)^3)}$, where the coefficients are determined by the positions of the on and off state attractors and are of the order of $\xi_1\approx0.01$ and $\xi_2\approx100$.  \\ 
Whether the switch is nonadiabatic or adiabatic can be determined by comparing the mean free path to the size of the transition region. If $l_{TST}/l_{mfp}>1$ many crossings are required and the transition is adiabatic. If $l_{TST}/l_{mfp}<1$ the system commits to the new attractor once it reaches the separatrix, hence the transition is nonadiabatic. In the strictly adiabatic regime the diffusion of the system is governed by protein diffusion induced by the birth-death process, as opposed to the weakly adiabatic regime, where diffusion due to churns dominates. A phase diagram showing the different escape mechanisms in parameter space for fixed $K^{eq} V^2$ is shown in Figure \ref{Diagram}. \\\
\textbf{Comparison with Numerically Exact Results}\\
While the mechanism of spontaneous switching or epigenetic escape is different in the various regimes, we understand the rates in all regimes using the notion of a stochastic separatrix. We can compare these approximations with numerical calculations due to Kepler and Elston \cite{KeplerElston, Gardner} and our own full numerical results. \\
In the nonadiabatic limit (small $\kappa=h g_{\uparrow}^2/(2 k^3)$) the escape process is determined by the rate of DNA state fluctuations. In this regime the rates are given by equations \ref{kon_0} and \ref{koff_0} (gray dashed line) (Figure \ref{Figure1}). These agree with the discrete numerical calculation of the mean free passage time from each basin. Our numerical calculations confirm that only in the extremely adiabatic limit (large $\kappa$ - flat escape rate) can the DNA fluctuations safely be neglected. Only for this extreme limit does the lifetime become determined by protein synthesis/ degradation fluctuations alone (light gray dashed line). Estimates of the input parameter would suggest that the weakly adiabatic regime is common for biological switches. In the weakly adiabatic regime the escape rate does not just depend on occupancy averaged growth rates, but still depends on the adiabaticity parameter, as shown in Figure \ref{Figure1}. Neglecting the influence of DNA fluctuations in this limit, as many treatments have done would give the extreme adiabatic asymptotic value of the escape rate also pictured on the graph. Both the strictly and weakly adiabatic regimes can be obtained from the more general calculation using the full diffusion coefficient. The full treatment is only required in a small crossover regime (gray solid line). \\
\textbf{Summary}\\
Spontaneous transitions between attractors of genetic systems are caused by coupled stochastic fluctuations in the DNA state and protein number. Even in parameter regimes where the DNA state locally would appear to reach a steady state much more rapidly than the protein number state, the fluctuations due to binding and unbinding of transcription factors greatly influence the protein number fluctuations and hence modify the rate of spontaneous transitions between epigenetic states. We call such a regime the weakly adiabatic by contrast to the strongly adiabatic limit, where the DNA binding state may be taken to be in equilibrium. The mechanism of spontaneous switching between stable attractors in the weakly adiabatic regime is graphically explained by a churning process, which causes protein numbers to fluctuate from the mean flow. How the escape rates $k_{on}$ and $k_{off}$ depend on molecular parameters in the nonadiabatic, weakly and strongly adiabatic should allow one to understand the evolutionary constraints necessary to achieve stable yet responsive switches, a topic we hope to return to. By considering both the DNA and protein degrees of freedom, the rate theories we have presented provide an intuitive description of spontaneous switching events, in terms of the molecular parameters that determine the functioning of a genetic switch.  

\begin{acknowledgements}
We acknowledge the supported by the Center for Theoretical Biological
Physics through National Science Foundation Grants PHY0216576 and
PHY0225630. We wish to thank Patrick H. Diamond and Jin Wang for insightful comments on the manuscript.
\end{acknowledgements}


\begin{thebibliography}{2}
\bibitem{SasaiWolynes} Sasai, M. \& Wolynes, P. G., (2003) \textit{Proc. Natl. Acad. Sci. USA} \textbf{100}, 2374-2379.
\bibitem{Kauffman} S.A. Kauffman, S. A., (1993) \textit{The Origins of Order} (Oxford University Press, London).
\bibitem{Ptashne} Ptashne, M., (2002)  \textit{Genes and Signals} (Cold Spring Harbor Laboratory Press, New York).
\bibitem{MonodSavageau} Jacob, F. \& Monod, J., (1961) \textit{J. Mol. Biol.}  \textbf{3}, 318-356.
\bibitem{Davidson} Davidson, E.H., Rast, J. P, Olivieri, P., Ransick, A., Calestani, C., Yuh, C. H., Minokawa, T., Amore, G., Hinman, V., Arenas-Mena, C., et al., (2002)  \textit{Science} \textbf{295}, 1669-1678.
\bibitem{ABOuden} Acar, M., Becskei, A., \& van Oudenaarden, A., (2005) \textit{Nature} \textbf{435}, 228-231.
\bibitem{BarYam} Huang, S., Eichler, G., Bar-Yam, Y. \& Ingber, D.E., (2005) \textit{Phys. Rev. Lett.} \textbf{94}, 128701-1-4.
\bibitem{AurellSneppen} Aurell, E., \& Sneppen, K., (2002) \textit{Phys. Rev. Lett.} \textbf{88}, 048101-1-4.
\bibitem{Little} Little, J. W., Shepley, D. P. \& Wert, D. W., (1999) \textit{ EMBO J.} \textbf{18}, 4299-4307.
\bibitem{WarrentenWolde} Warren, P. B. \& ten Wolde, P. R., (2005) \textit{J. Phys. Chem. B} \textbf{109}, 6812-6823.
\bibitem{Bialek} Bialek, W., (2001) \textit{Advances in Neural Information Processing 13} eds. Leen, T. K., Dietterich, T. G., \& Tresp,V. (MIT Press, Cambridge), pp. 159-165. 
\bibitem{MetzlerWolynes} Metzler, R. \& Wolynes, P. G., (2002) \textit{Chem. Phys.}, \textbf{284}, 469-479. 
\bibitem{OnuchicWolynes88} J.N. Onuchic and Wolynes, P. G., (1988) \textit{Journal Phys. Chem.}, \textbf{92}, 6495-6503.
\bibitem{WolynesSF} Wolynes, P. G., (1989) in \textit{Lectures in Science and Complexity, SFI Studies in the Sciences of Complexity}, ed. Stein, D. (Addison-Wesley Longman), 355-387. 
\bibitem{Wigner} Wigner, E., (1932) \textit{Phys. Rev.}, \textbf{40}, 749-759.
\bibitem{FrauenfelderWolynes} Frauenfelder H. \& Wolynes, P. G., (1985) \textit{Science} \textbf{229}, 337-345.
\bibitem{Pedraza} Pedraza, J. M. \& van Oudenaarden, A., (2005) \textit{Science} \textbf{307}, 1965-1969.
\bibitem{Paulsson} Paulsson, J., (2004) \textit{Nature} \textbf{427}, 415-418.
\bibitem{Walczak} Walczak, A. M., Sasai, M. \& Wolynes, P. G., (2005) \textit{Biophys. J.} \textbf{88}, 828-850.
\bibitem{Hornos} Hornos, J. E. M., Schultz, D., Innocentini, G. C. P., Wang, J., Walczak, A. M., Onuchic, J. N. \& Wolynes, P. G., (2005) \textit{Phys. Rev. E} \textbf{72}, in press. 
\bibitem{SteinMaier} Maier, R. S. \& Stein, D. L., (1993) \textit{Phys. Rev. E} \textbf{48}, 931-938.
\bibitem{KeplerElston} Kepler, T. B. \& Elston, T. C., (2001) \textit{Biophys. J.} \textbf{81}, 3116-3136.
\bibitem{Romaetal} Roma, D. M, O'Flanagan, R. A., Ruckenstein, A. E., Sengupta, A. M. \& Mukhopadhyay, R., (2005) \textit{Phys. Rev. E} \textbf{71}, 011902-1-5.
\bibitem{AllenWarrentenWolde} Allen, R. J., Warren, P. B. \& ten Wolde, P. R., (2005) \textit{Phys. Rev. Lett.} \textbf{94}, 018104-1-4.
\bibitem{Hwa} Buchler, N. E., Gerland, U. \& Hwa, T., (2003) \textit{Proc. Natl. Acad. Sci. USA} \textbf{100}, 5136-5141.
\bibitem{OzbudakOuden} E.M Ozbudak, E. M., Thattai, M., Lim, H.N., Shraiman, B.I. \& van Oudenaarden, A., (2004) \textit{Nature} \textbf{427}, 737-740.
\bibitem{WOW} Wang, J., Onuchic, J., \& Wolynes, P. G., (1996) \textit{Phys. Rev. Lett.} \textbf{76}, 4861-4865.
\bibitem{Ninio} Ninio, J., (1987) \textit{Proc. Natl. Acad. Sci. USA} \textbf{84}, 663-667. 
\bibitem{AurellSneppenPRE} Aurell, E., Brown, S., Johanson, J., \& Sneppen, K., (2002) \textit{Phys. Rev. E} \textbf{65}, 051914-1-9. 
\bibitem{Gardner} Gardiner, C. W., (1990) \textit{Handbook of Stochastic Methods} (Springer-Verlag, Berlin).
\end{thebibliography}
\end{document}